\documentclass[twocolumn,showpacs,showkeys,superscriptaddress]{revtex4}

\usepackage{amsmath}
\usepackage{bbold}
\usepackage{babel}
\usepackage{amsfonts} 
\usepackage{amssymb}
\usepackage{pbsi}
\usepackage[T1]{fontenc}
\usepackage{hyperref}
\usepackage{xcolor}
\usepackage{graphicx}
\usepackage{subfig}
\usepackage{float}
\usepackage{soul}

\usepackage{color}

\begin{document}

\title{Accelerating solutions of the Korteweg-de Vries equation}

\author{Maricarmen A. Winkler}
\email{ma.winkler822@gmail.com}
\affiliation{Facultad de Ingenier\'ia y Ciencias,
Universidad Adolfo Ib\'a\~nez, Santiago 7491169, Chile.}

\author{Felipe A. Asenjo}
\email{felipe.asenjo@uai.cl}\affiliation{Facultad de Ingenier\'ia y Ciencias,
Universidad Adolfo Ib\'a\~nez, Santiago 7491169, Chile.}

\begin{abstract}
The Korteweg-de Vries  equation is a fundamental nonlinear equation that  describes solitons with constant velocity. On the contrary, here we show that this equation also presents accelerated wavepacket solutions. This behavior is achieved by putting the Korteweg-de Vries  equation in terms of the Painlev\'e I equation. The accelerated waveform solutions are explored numerically showing their accelerated behavior explicitly.

\end{abstract}


\maketitle

\section{Introduction}

The Korteweg-de Vries (KdV) equation is widely recognized as a fundamental tool for describing nonlinear wave phenomena across various scientific disciplines \cite{whitham1974linear,drazin1989solitons,scott2003nonlinear,remoissenet2013waves}, with the solitary wave, or soliton, being one of its notable closed-form solutions. First derived in 1895 \cite{korteweg1895change}, the KdV equation was formulated to model the propagation of surface water waves, demonstrating that the soliton represents the displacement of the water surface from its equilibrium.

Solitons are localized structures that arise in systems where nonlinearity and dispersion are delicately balanced. Due to this equilibrium, solitons are remarkably stable, maintaining their shape and velocity even after interactions with other solitons. The observation of these nonlinear waveforms in a variety of natural contexts has led to significant theoretical developments, with applications spanning fields such as optics \cite{Alhami2022,He2014,Horsley2016}, fluid dynamics \cite{BOUSSINESQ1877,Tian2001}, conducting polymers \cite{NARIBOLI1970661,Crighton1995}, gravitation \cite{suleimanov1994onset,lidsey2012cosmology}, and biology \cite{bilotta2013cellular,carstea2015coupled}, among others.
For instance, in fluid dynamics, its original context, the KdV equation is used for studying ocean waves, tidal bores, and internal waves in stratified fluids. 
In plasma physics, is very common to use the KdV equation to describe the dynamics of ion-acoustic waves, magnetoacoustic waves and Alfvén waves in inhomogeneous plasmas, such as a dusty plasma \cite{tran1974korteweg,tariq2023backlund,mowafy2008effect,zhang2023application}.
In nonlinear optics, the KdV equation helps the understanding of pulse propagation in optical fibers and waveguides \cite{rodriguez2003standard}. These optical solitons are very important for modern telecommunications, where they help mitigate signal degradation.
The KdV equation also finds applications in biological systems, particularly in the study of nerve pulse propagation \cite{fongang2018breathing}. 

The general form of the KdV equation is mathematically represented as
\begin{equation}
    \frac{\partial \phi}{\partial t} + p \phi \frac{\partial \phi}{\partial \xi} + q \frac{\partial^3 \phi}{\partial \xi^3} = 0\ , \label{kdv} 
\end{equation}
where $\phi(t, \xi)$ is the wave profile as a function of time $t$ and a position-like coordinate $\xi$. Besides, $p$ is the nonlinear coefficient and $q$ the dispersion coefficient. 

It is well-known that Eq.~\eqref{kdv} is integrable allowing for exact solutions, including the soliton solution. These solitons have constant velocity propagation (no acceleration), as it is summarized in Sec.~\ref{soliton}.
However, we show here that other propagating solutions, with not constant velocity, are possible to be found. The aim of this work is to investigate and discuss a new kind of accelerated solution of the KdV equation, which differs from the conventional soliton solution, in its form of propagation as well as its wave profile. 
This analytically accelerated wave solution, presented in Sec.~\ref{accelerated}, is formed by a wavepacket that satisfies a Painlev\'e I equation, and that mainly propagates along the accelerated coordinate, depending on $t^2$. In Sec.~\ref{numerical}, the behavior of this solution is numerically studied, to finish with a discussion on the implications of this solution in Sec.~\ref{conclusions}.

\section{Soliton solution}\label{soliton}

Typically, the soliton solution of Eq.~\eqref{kdv} has the form \cite{WAZWAZ2008485}
\begin{equation}
    \phi\left(t, \xi\right)=\phi_0\  \text{sech}^2\left(\frac{\eta}{\Delta}\right)\ , \label{sol1}
\end{equation}
\newline where $\eta=\xi-vt$ is the  coordinate solidary to the soliton propagation, with constant  $v$, representing a velocity-like quantity. Besides, $\phi_0=3v/p$ is the amplitude of the soliton that depends on the nonlinear coefficient $p$. Also, $\Delta=\sqrt{4|q|/v}$ represents the soliton width, depending on the dispersion coefficient $q>0$. Notice that this a travelling soliton solution, which does not exist for $v=0$. 

The soliton \eqref{sol1}  travels with constant velocity, as it can be seen by considering the coordinate frame $\eta=\mbox{constant}$, which implies $d\xi/dt=v$. This an inherent feature of  solitons satisfying Eq.~\eqref{kdv}.

On the other hand, when $q<0$, the travelling soliton solution becomes \cite{fan2003new,winkler2024existence}  
\begin{equation}
    \phi\left(t, \xi\right)=\frac{\phi_0}{2} \text{tanh}^2\left(\frac{\eta}{\Delta\sqrt{2}}\right)\ , \label{sol2}
\end{equation}
where the amplitude of the soliton is reduced in half and its width is increased by $\sqrt{2}$. This soliton also travels with constant velocity.

Both soliton solutions \eqref{sol1} and \eqref{sol2} have opposite behaviors. When the soliton \eqref{sol1} is compressive (rarefactive), the soliton \eqref{sol2} is rarefactive (compressive). However, both propagate with the same constant velocity.

\section{Accelerating wavepacket}\label{accelerated}

Contrary to the previous section, 
here we show that  Eq.~\eqref{kdv} has also accelerated solutions. This means that the wave profile $\phi$ will depend on a combination of coordinates that make the velocity change, such as  $d\xi/dt\neq \mbox{constant}$. As we will see below, this implies that constant accelerated propagation is only possible for a wavepacket,  not for a soliton.

Let us consider the ansazt for the waveform as
\begin{equation}
    \phi\left(t, \xi\right)=a\, \gamma\,  \varphi (x)+\frac{a}{p} t\, ,
    \label{newamp}
\end{equation}
depending on the accelerated coordinate 
\begin{equation}
    x=\alpha\left(\xi - \frac{1}{2}at^2\right)\, .  \label{coordenada}
\end{equation}
Here, $\gamma$ and $\alpha$ are constants to be determined, and $a$  can be understood as the acceleration of propagation of the wavepacket. The function $\varphi$ determines the profile of the accelerated wavepacket.
Notice that the form of the ansatz has been chosen such that it identically vanishes when there is no acceleration, $a=0$.

Using the profile \eqref{newamp}, Eq.~\eqref{kdv} can be written for $\varphi$ in terms of coordinate $x$ as
\begin{align}
    \frac{\partial^3 \varphi}{\partial x^3} + \lambda\frac{\partial \varphi^2}{\partial x} + \delta &= 0 \label{eq2}
\end{align}
where $\lambda=a p\gamma/(2q\alpha^2)$, and $\delta= 1/(pq\alpha^3\gamma)$. Both parameters quantify the relation between acceleration, the nonlinear coefficient and the dispersion coefficient.

A closer look into Eq.~\eqref{eq2} reveals that it can be integrated to
\begin{align}
       \frac{\partial^2 \varphi}{\partial x^2}+\lambda \varphi^2+\delta x &= 0\ , \label{eq3}
\end{align}
where we  have set the integration constant to be $0$.
From here, we notice that when $\lambda=-6$ and $\delta=-1$, then Eq.~\eqref{eq3} turns into a Painlevé I equation
\begin{equation}
    \frac{\partial^2 \varphi}{\partial x^2}=6\varphi^2+x\, ,
    \label{painleve}
\end{equation}
which is achieved under the specific choices for
\begin{eqnarray}
    \alpha&=&\left(\frac{a}{12 q^2} \right)^{1/5}\, ,\label{alpha}\\
    \gamma&=&-\frac{1}{pq}\left(\frac{a}{12 q^2} \right)^{-3/5}\, . \label{gamma}
\end{eqnarray}

Eq.~\eqref{painleve} is 
the main result of this work. It shows that the KdV equation can be put in terms of the Painlevé I equation for a wavepacket with accelerated propagation along coordinate \eqref{coordenada},  implying that $d\xi/dt=a t$. This produces a typical accelerated (parabolic) behavior in $\xi$-$t$ space.

Furthermore, the dynamics of the solution is solely determined by the constant parameters $p$, $q$, and $a$, the same number of  parameter than the soliton solution ($p$, $q$, and $v$). In order to properly show the impact of the acceleration parameter $a$ on the wavepacket propagation, Eq.~\eqref{painleve} can be solved numerically. In general,  solutions
of Eq.~\eqref{painleve} do not represent a solitary wave.

\section{Numerical solution}\label{numerical}

Eq.~\eqref{painleve} is 
the first of six equations, where their solutions are the Painlevé transcendents \cite{davis1960introduction,Ablowitz_Clarkson_1991,CLARKSON2003}. They have been highly studied in the context of dispersive long waves, pumped Maxwell-Bloch systems and physical problems leading to second order ordinary differential equations \cite{porsezian1995optical,feng2017nonlocal,levi2013painleve}

\begin{figure}[h]\centering
\includegraphics[width=7.3cm]{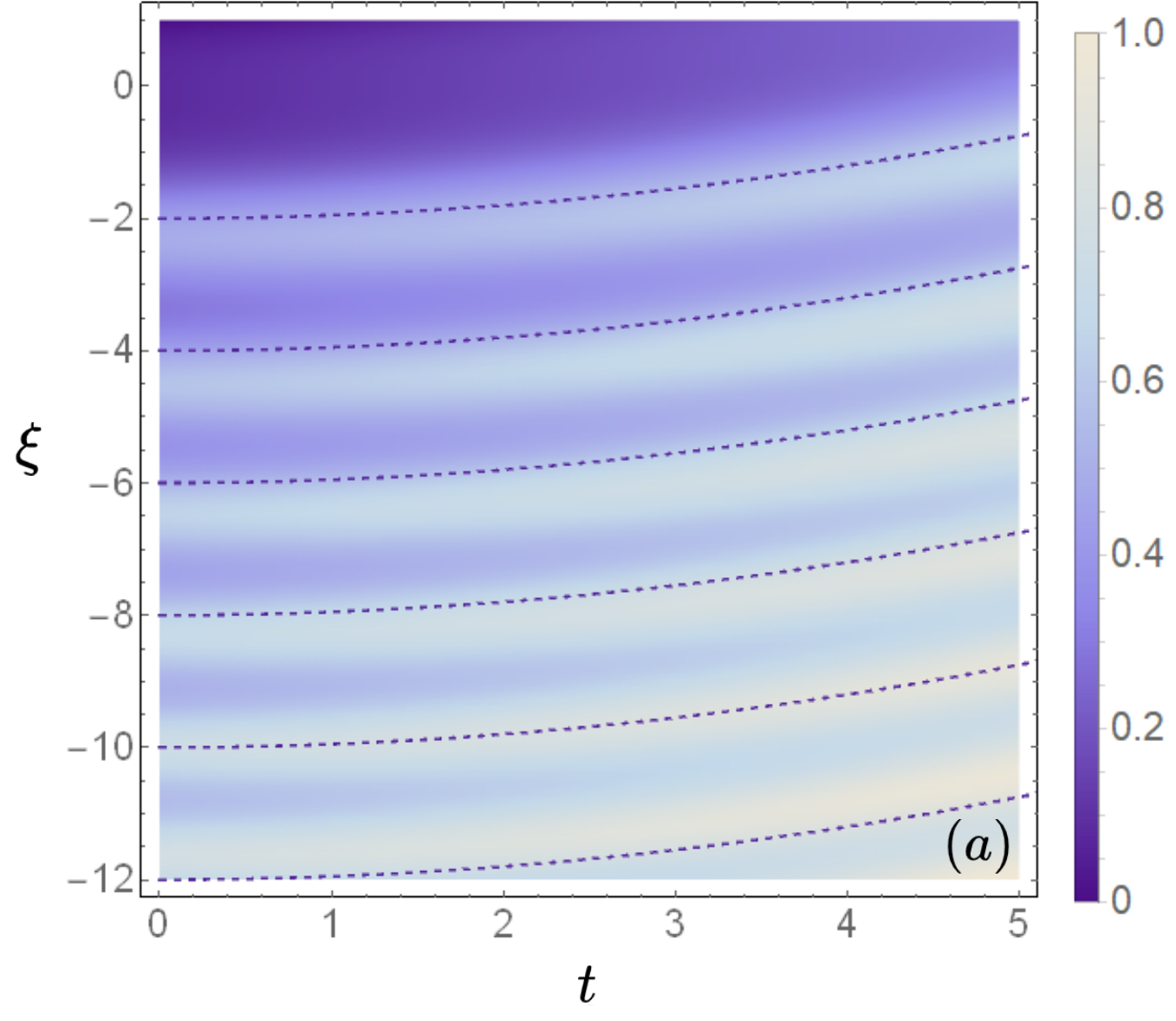}
\includegraphics[width=7.3cm]{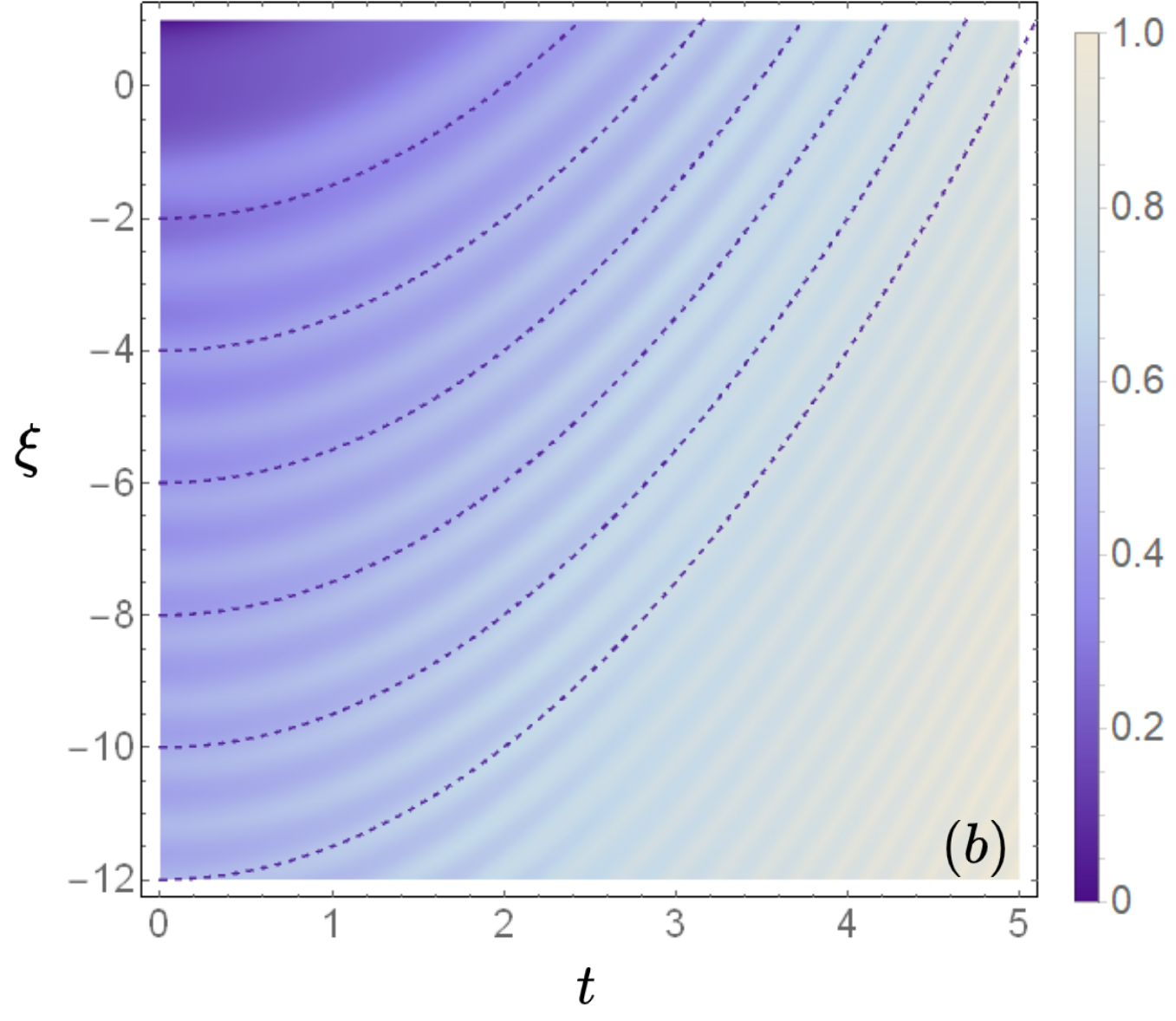}
\caption{Density plot for solution $\phi(t,\xi)$ [Eq.~\eqref{newamp}], with  $p=20$ and $q=0.1$, and initial conditions $\varphi(x = 0) = 0.1$ and $d_x\varphi(x = 0) = 0$. Dashed lines represent parabolic trajectories for (a) $a=0.1$, and  (b) $a=1$.} \label{fig1}
\end{figure}

\begin{figure}[h]\centering
\includegraphics[width=7.3cm]{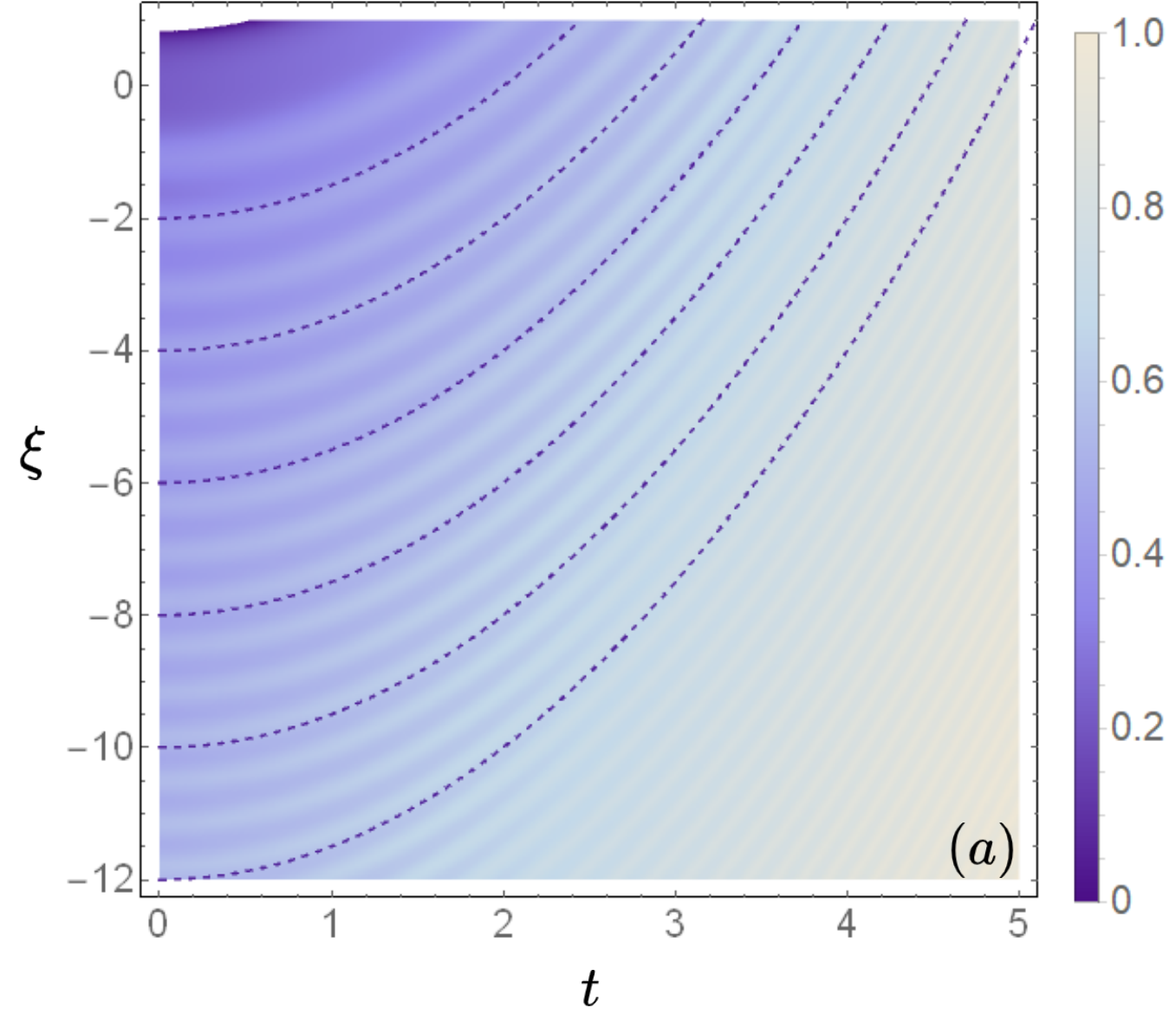}
\includegraphics[width=7.3cm]{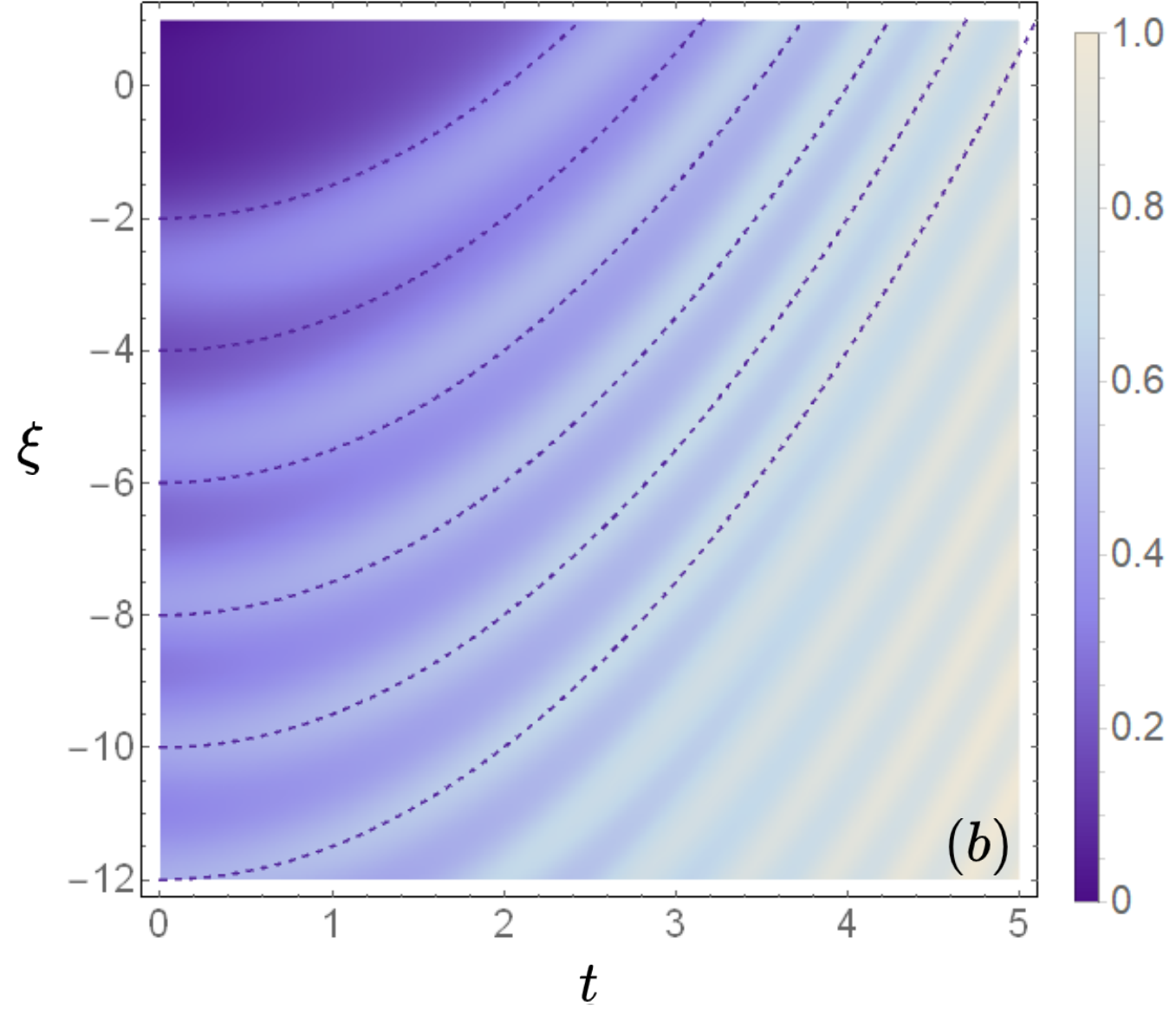}
\caption{Density plot for $\phi(t,\xi)$ [Eq.~\eqref{newamp}], with
$p=20$ and $a=1$, for initial conditions
$\varphi(x = 0) = 0.1$ and $d_x\varphi(x = 0) = 0$. Dashed lines represent parabolic trajectories for (a) $q=0.05$ and  (b) $q=0.5$.}\label{fig2}
\end{figure}

In order to evaluate the accelerated solution
\eqref{newamp}, we solve numerically Eq.~\eqref{painleve}. This is shown in Figs.~\ref{fig1} and \ref{fig2}, in terms of density plots. 
To find this solution, we set values for the nonlinear coefficient $p$, the dispersion coefficient $q$ and the acceleration-like parameter $a$,  that make the solution a real function. 
To solve Eq.~\eqref{painleve}, we chose the initial conditions $\varphi (x=0)= 0.1$ and $d_x\varphi (x=0)= 0$.  In general, we see that each part of the wavepacket follows accelerated (parabolic) trajectories, represented by the dashed lines on both figures. These curves are determined by the constant value of the accelerated coordinate \eqref{coordenada} (defining different initial conditions). Therefore, all the curved trajectories satisfy $d^2\xi/dt^2=a$. In both figures, for the curves, we have chose different initial positions at $x= -2, -4, -6, -8, -10, -12$.

In Fig.~\ref{fig1} we explore two different values for the acceleration parameter $a$. The parabolic behaviour of the wave train \eqref{newamp}  is altered by the value of this parameter, increasing the curvature of the trajectory for higher values of $a$. In Fig. \ref{fig1}(a) we show the accelerated solution \eqref{newamp} for $a=0.1$. In this case, the wavepacket has several minima and maxima, all following the accelerated coordinate \eqref{coordenada}. In Fig. \ref{fig1}(b) we change the value of acceleration to $a=1$, where the increase in the curvature of the trajectory  that the wave follows can be easily seen for the same initial conditions. The increase on curvature of the trajectories is a direct consequence of the larger acceleration of the wavepacket.
In both cases we use fixed values for the $p$ and $q$ coefficients, demonstrating the sole effect of acceleration. 

According to Eqs. \eqref{alpha} and \eqref{gamma}, the nonlinear and dispersion coefficients are free parameters that  modified the characteristics of the  wave train, for a given acceleration. This is depicted in Fig.~\ref{fig2}.
When the parameter $q$ is allowed to change, we can observe how dispersion changes for the accelerated solution. For low values of $q$, the wavepacket has more minima and maxima than in the case shown in Fig.~\ref{fig1}, and they seem to appear less often than with higher values of $q$. This is the expected behavior of dispersion.
This can be seen in Figs.~\ref{fig2}(a) and \ref{fig2}(b), where we have used $q=0.05$ and $q=0.5$, respectively. In both cases is shown by dashed lines that modifying the dispersion coefficient does not change the trajectory of the wave packet. This is because the
accelerated coordinate \eqref{coordenada} does not depend on $q$. Thus, for different values of $q$, the parabolic trajectories remain unchanged for different initial conditions.

\section{Summary}\label{conclusions}

In this study we demonstrate that the Korteweg-de Vries equation \eqref{kdv}  also have a set of accelerating wavepacket solutions, distinct from solitons. This result expands the understanding of nonlinear wave dynamics to a new class of unexplored solutions.

This accelerated wave train is obtained by reducing the KdV equation into a Painlevé I equation, in terms of an accelerated coordinate system. Thus, the new solution for the KdV equation  describes a wavepacket following a parabolic trajectory in $\xi$-$t$ space, and it is susceptible to changes due to the different parameters, 
specifically the acceleration $a$ and the dispersion coefficient $q$. 

This  solution for the KdV equation belongs to a new class of accelerated solutions for nonlinear dynamics systems, such as the ones described by the nonlinear Schr\"odinger equation \cite{Asenjonon}. Thus, we think that the presented results contribute to a novel understanding on 
the KdV equation's versatility, and its relevance in modern theoretical and applied physics.

\section*{Acknowledgments}
The authors of this work thank to
FONDECYT postdoc grant No. 3240441 (MAW), and to
FONDECYT grant No. 1230094 (FAA) for their support.

\bibliographystyle{h-physrev}

\bibliography{unnamed}

\begin{thebibliography}{10}

\bibitem{whitham1974linear}
G.~Whitham,
\newblock {\itshape Linear and Nonlinear Waves} (Wiley, 1974).

\bibitem{drazin1989solitons}
P.~Drazin and R.~Johnson,
\newblock {\itshape Solitons: An Introduction} (Cambridge University Press, 1989).

\bibitem{scott2003nonlinear}
A.~Scott,
\newblock {\itshape Nonlinear Science: Emergence and Dynamics of Coherent Structures}, 2nd ed. (Oxford University Press, 2003).

\bibitem{remoissenet2013waves}
M.~Remoissenet,
\newblock {\itshape Waves Called Solitons: Concepts and Experiments} (Springer Science \& Business Media, 2013).

\bibitem{korteweg1895change}
D.~Korteweg and G.~de~Vries,
\newblock Phyl.\ Mag. {\bfseries 39}, 422 (1895).

\bibitem{Alhami2022}
R.~Alhami and M.~Alquran,
\newblock Opt.\ Quantum Electron. {\bfseries 54}, 553 (2022).

\bibitem{He2014}
J.~He, L.~Wang, L.~Li, K.~Porsezian, and R.~Erd\'elyi,
\newblock Phys.\ Rev.\ E {\bfseries 89}, 062917 (2014).

\bibitem{Horsley2016}
S.~A.~R. Horsley,
\newblock J.\ Opt. {\bfseries 18}, 085104 (2016).

\bibitem{BOUSSINESQ1877}
J.~Boussinesq,
\newblock Bibliothèque nationale de France {\bfseries 1 vol. (XXII-680-61 p.)}, 680 (1877).

\bibitem{Tian2001}
B.~Tian and Y.-T. Gao,
\newblock Eur.\ Phys.\ J.\ B {\bfseries 22}, 351 (2001).

\bibitem{NARIBOLI1970661}
G.~Nariboli and A.~Sedov,
\newblock J.\ Math.\ Anal.\ Appl. {\bfseries 32}, 661 (1970).

\bibitem{Crighton1995}
D.~G. Crighton,
\newblock Acta Appl.\ Math. {\bfseries 39}, 39 (1995).

\bibitem{suleimanov1994onset}
B.~Suleimanov,
\newblock Zhurn. Eskper. Teor. Fiz {\bfseries 105}, 1089 (1994).

\bibitem{lidsey2012cosmology}
J.~E. Lidsey,
\newblock Phys.\ Rev.\ D {\bfseries 86}, 123523 (2012).

\bibitem{bilotta2013cellular}
E.~Bilotta and P.~Pantano,
\newblock Int.\ J.\ Bifurcat.\ Chaos {\bfseries 23}, 1330003 (2013).

\bibitem{carstea2015coupled}
A.~Carstea and T.~Tokihiro,
\newblock J.\ Phys.\ A {\bfseries 48}, 055205 (2015).

\bibitem{tran1974korteweg}
M.~Q. Tran and P.~J. Hirt,
\newblock Plasma Physics {\bfseries 16}, 617 (1974).

\bibitem{tariq2023backlund}
M.~S. Tariq {\em et~al.},
\newblock Phys.\ Fluids {\bfseries 35} (2023).

\bibitem{mowafy2008effect}
A.~Mowafy, E.~El-Shewy, W.~Moslem, and M.~Zahran,
\newblock Phys.\ Plasmas {\bfseries 15} (2008).

\bibitem{zhang2023application}
H.~Zhang {\em et~al.},
\newblock J.\ Plasma Phys. {\bfseries 89}, 905890212 (2023).

\bibitem{rodriguez2003standard}
R.~Rodr{\'\i}guez, J.~Reyes, A.~Espinosa-Cer{\'o}n, J.~Fujioka, and B.~Malomed,
\newblock Phys.\ Rev.\ E {\bfseries 68}, 036606 (2003).

\bibitem{fongang2018breathing}
G.~Fongang~Achu, F.~Moukam~Kakmeni, and A.~Dikande,
\newblock Phys.\ Rev.\ E {\bfseries 97}, 012211 (2018).

\bibitem{WAZWAZ2008485}
A.-M. Wazwaz,
\newblock Chapter 9 the kdv equation,
\newblock , Handbook of Differential Equations: Evolutionary Equations Vol.~4, pp. 485--568, North-Holland, 2008.

\bibitem{fan2003new}
E.~Fan,
\newblock Chaos, Solitons \& Fractals {\bfseries 15}, 567 (2003).

\bibitem{winkler2024existence}
M.~A. Winkler, V.~Mu{\~n}oz, and F.~A. Asenjo,
\newblock Fundamental Plasma Physics {\bfseries 9}, 100030 (2024).

\bibitem{davis1960introduction}
H.~Davis and U.~A.~E. Commission,
\newblock {\itshape Introduction to Nonlinear Differential and Integral Equations} (U.S. Atomic Energy Commission, 1960).

\bibitem{Ablowitz_Clarkson_1991}
M.~A. Ablowitz and P.~A. Clarkson,
\newblock {\itshape Solitons, Nonlinear Evolution Equations and Inverse Scattering}London Mathematical Society Lecture Note Series (Cambridge University Press, 1991).

\bibitem{CLARKSON2003}
P.~A. Clarkson,
\newblock J. Comput. Appl. Math. {\bfseries 153}, 127 (2003),
\newblock Proceedings of the 6th International Symposium on Orthogonal Poly nomials, Special Functions and their Applications, Rome, Italy, 18-22 June 2001.

\bibitem{porsezian1995optical}
K.~Porsezian and K.~Nakkeeran,
\newblock J.\ Mod. Opt. {\bfseries 42}, 1953 (1995).

\bibitem{feng2017nonlocal}
L.-L. Feng, S.-F. Tian, and T.-T. Zhang,
\newblock Z.\ Naturforsch {\bfseries 72}, 425 (2017).

\bibitem{levi2013painleve}

\newblock D.~Levi and P.~Winternitz{\itshape Painlev{\'e} transcendents: their asymptotics and physical applications} Vol. 278 (Springer Science \& Business Media, 2013).

\bibitem{Asenjonon}
F.~A. Asenjo,
\newblock J.\ Plasma Phys. {\bfseries 90}, 905900119 (2024).

\end{thebibliography}

\end{document}